\documentclass{PoS}
\usepackage{amsmath}
\usepackage{bm}
\usepackage{amsbsy}

\def\nab#1{{\nabla_{#1}}}
\def\nabstar#1{{\nabla\kern0.5pt\smash{\raise 4.5pt\hbox{$\ast$}}
               \kern-5.5pt_{#1}}}

\def\nabbar#1{{\overset{\leftarrow}{\nabla}_{#1}}}
\def\nabbarstar#1{{\overset{\leftarrow}{\nabla}\kern0.5pt\smash{\raise 4.5pt\hbox{$\ast$}}
               \kern-5.5pt_{#1}}}

\def\nabdbar#1{{\overset{\leftrightarrow}{\nabla}_{#1}}}
\def\nabdbarstar#1{{\overset{\leftrightarrow}{\nabla}\kern0.5pt\smash{\raise 4.5pt\hbox{$\ast$}}
               \kern-5.5pt_{#1}}}
\def\psibar{\overline{\psi}}
           
\title{The energy-momentum tensor in lattice QCD and the Equation of State}

\ShortTitle{The energy-momentum tensor in lattice QCD and the Equation of State}

\author{Mattia Dalla Brida\\
         Dipartimento di Fisica, Universit\`a di Milano-Bicocca\\
         and INFN, sezione di Milano-Bicocca\\
         Edificio U2, Piazza della Scienza 3\\ 
         20126 Milano, Italy.\\
         E-mail: \email{mattia.dallabrida@unimib.it}}

\author{Leonardo Giusti\\
         Dipartimento di Fisica, Universit\`a di Milano-Bicocca\\
         and INFN, sezione di Milano-Bicocca\\
         Edificio U2, Piazza della Scienza 3\\ 
         20126 Milano, Italy.\\
         E-mail: \email{leonardo.giusti@unimib.it}}

\author{\speaker{Michele Pepe}\\
        INFN, Sezione di Milano-Bicocca\\ 
        Edificio U2, Piazza della Scienza 3\\ 
        20126 Milano, Italy.\\
        E-mail: \email{michele.pepe@mib.infn.it}}

\abstract{We present a new theoretical and practical strategy to renormalize non-perturbatively the energy-momentum tensor in lattice QCD
  based on the framework of shifted boundary conditions. As a preparatory step for the fully non-perturbative calculation, we apply the
  strategy at 1-loop order in perturbation theory determining the renormalization constants of both the gluonic and the fermionic
  components of the energy-momentum tensor. Using shifted boundary conditions, the entropy density of QCD is directly related to the
  expectation value of the space-time components of the renormalized energy-momentum tensor. We then discuss its practical implementation by
  numerical simulations of QCD with 3 flavours of Wilson quarks for temperatures between 2.5~GeV and 80~GeV.}

\FullConference{37th International Symposium on Lattice Field Theory - Lattice2019\\
		16-22 June 2019\\
		Wuhan, China}

\begin{document}

\section{Introduction}

The energy-momentum tensor (EMT) is a fundamental quantity of a Quantum Field Theory since it contains the currents associated to Poincar\'e
symmetry and scale invariance. Furthermore, the expectation values of its matrix elements are physical quantities: they are related to thermal
features like the pressure, the entropy density and the energy density as well as to transport coefficients. In the
continuum, the EMT is a conserved quantity and it does not renormalize; the formulation of the EMT on the lattice was
investigated by Caracciolo et al.~\cite{Caracciolo:1989pt} who showed that the breaking of translation and rotation symmetries by the lattice
regularization  leads to a non trivial definition of the EMT on the lattice. The $\{ 10 \}$-dimensional symmetric representation of the EMT
in the continuum reduces to the sum of the $\{ 6 \}$, the $\{ 3 \}$ and the $\{ 1 \}$ representations of the hypercubic group of the lattice
with finite renormalization constants that approach 1 in the continuum limit. Imposing the validity of Ward Identities associated to Poincar\'e
symmetry, they defined the renormalization constants perturbatively, computing them at 1-loop order~\cite{Caracciolo:1991cp}.

However, it was not clear how to define non-perturbatively the renormalization constants and, consequently, the EMT itself. Using shifted
boundary conditions, in~\cite{Giusti:2010bb, Giusti:2011kt, Giusti:2012yj} it was proposed a non-perturbative definition of the EMT for the
$SU(N_c)$ Yang-Mills theory. In that framework one can exploit new Ward Identities that, in the case of periodic boundary conditions, become
trivial identities. It is important to note that the non-perturbative definition of the renormalization constants with shifted
boundary conditions provided also a practical method for their numerical computation by Monte Carlo simulations on the
lattice. In~\cite{Giusti:2015daa}, the renormalization constants of the EMT for the $SU(3)$ Yang-Mills theory have been computed
non-perturbatively for the first time.  

In these proceedings we extend the above results to the case of QCD. We consider the Wilson formulation of quarks on the lattice
and the Sheikoleslami-Wohlert improved action. In the case of QCD, the EMT has gluonic
and fermionic parts that mix under renormalization; in order to disentangle those two components, it is useful to introduce a twist phase for the matter fields in the temporal boundary conditions: it can also be interpreted as an imaginary baryonic chemical potential. Similar to the
pure gauge case, the setup of shifted boundary conditions not only allows a non-perturbative definition of the EMT for QCD on the
lattice but it also represents again a practical method for the numerical computation by Monte Carlo simulations.

In the framework of shifted boundary conditions, the expectation values of the space-time matrix elements of the EMT do not vanish and they are
related by a purely geometric multiplicative factor to the entropy density of the system. That represents an alternative method to
investigate the Equation of State of a Quantum Field Theory with respect to the standard technique~\cite{Boyd:1996bx} based on the
measurement of the trace anomaly. Moreover, at variance with this latter method, no ultraviolet power divergence needs to be subtracted and
the continuum limit of the entropy density can be attained in a very simple way. This approach has turned out to be very successful in the
case of the $SU(3)$ Yang-Mills theory where it has been used for the calculation of the Equation of State with a permille accuracy up to the
temperature of 230 $T_c$, where $T_c$ is the critical temperature~\cite{Giusti:2014ila, Giusti:2016iqr}. The same method can be used to
compute the Equation of State of QCD and we discuss its implementation by numerical simulations with 3 flavours of Wilson quarks for
temperatures between 2.5~GeV and 80~GeV, i.e. the electro-weak scale. Preliminary results have been 
presented~in~\cite{DallaBrida:2017sxr,DallaBrida:2019gdt}. Finally, another approach based on the Gradient
Flow~\cite {Luscher:2010iy} has been also recently explored~\cite{Makino:2014taa,Taniguchi:2016ofw}.

\section{Lattice QCD with shifted boundary conditions}
We consider a four-dimensional lattice with temporal extent $L_0$ and size $L$ along the three spatial directions and we set the lattice
spacing $a = 1$. The gauge field is represented by the link variables $U_\mu(x)$ and the quark and anti-quark fields are given by $\psi(x)$
and $\psibar (x)$, respectively. The QCD action is $S=S^G +S^F$ where the gluonic part, $S^G$, and the fermionic part, $S^F$, are given by
\begin{equation}
  S^G={1\over g_0^2} \sum_x \sum_{\mu,\nu=0}^3 \mbox{Re Tr} [ 1-U_{\mu\nu} (x) ],
  \qquad
  S^F = \sum_x \psibar (x) (D+M_0) \psi(x)
\end{equation}
where $g_0$ is the bare coupling, $U_{\mu\nu}=U_\mu (x) U_\nu (x+\hat\mu)  U_\mu^\dagger (x+\hat\nu)  U_\nu^\dagger (x) $ is the
plaquette field and $M_0$ is the bare quark mass matrix. We consider the $O(a)$-improved Wilson-Dirac operator $D=D_{\rm w}+D_{\rm sw}$
given by the sum of the massless Wilson-Dirac operator, $D_{\rm w}$, and of the Sheikholeslami-Wohlert (SW)
term, $D_{\rm sw}$~\cite{Sheikholeslami:1985ij}.
The gauge field is shifted by a spatial vector ${\boldsymbol{\xi}}$ when crossing the 
boundary in the temporal direction and it is periodic in the spatial directions
\begin{equation}
  U_\mu (x_0+L_0,{\bf x}) = U_\mu (x_0,{\bf x}-L_0 {\boldsymbol\xi}),
  \qquad
  U_\mu (x_0, {\bf x}+L \hat k) = U_\mu (x_0,{\bf x}).
\end{equation}
In a similar way, the fermionic fields satisfy periodic boundary conditions in space and shifted boundary conditions in time in addition to the
usual anti-periodicity 
\begin{equation}
  \label{eq:psibcs}
  \psi(x_0+L_0,{\bf x}) = -e^{i \theta} \psi(x_0, {\bf x} - L_0 {\boldsymbol \xi}) 
  \qquad \mbox{and} \qquad
  \psi(x_0, {\bf x}+L \hat{k} ) =\psi(x_0, {\boldsymbol x}),
\end{equation}
\begin{equation}      
  \label{eq:psibarbcs}
  \psibar(x_0+L_0, {\bf x})=- e^{-i \theta} \psibar(x_0, {\bf x} - L_0 {\boldsymbol \xi})
  \qquad \mbox{and} \qquad
  \psibar(x_0, {\bf x} + L \hat{k} )= \psibar(x_0, {\bf x}),
\end{equation}
where we have included also a non-trivial phase at the temporal boundary. It is important to note that, by a redefinition of the fermionic fields,
the multiplicative phase $e^{\pm i\theta}$ can be removed from the temporal boundary conditions and it shows up as an imaginary baryonic
chemical potential. The partition function of the system and the free-energy density are given by
\begin{equation}\label{partfunct}
Z[L_0,L;{\boldsymbol \xi},\theta]= \int \mathcal{D}U\mathcal{D}\psi\mathcal{D}\psibar \,e^{-S},
\qquad\qquad
f[L_0,L; {\boldsymbol \xi},\theta] = -{1\over L_0L^3} \ln Z[L_0,L; {\boldsymbol \xi},\theta].
\end{equation}
The free-energy density provides a clear physical interpretation of the shift since in the thermodynamic limit it holds that
$f(L_0; {\boldsymbol \xi},\theta) = f({L_0 \sqrt{1+{\boldsymbol \xi}^2}}; {\bf 0}, \theta)$: this equation is the generalization to QCD of a
similar relation holding for the Yang-Mills theory~\cite{Giusti:2012yj}.\\ 
The EMT is given by the sum of a gluonic part and a fermionic one, $ T_{\mu\nu}= T_{\mu\nu}^G+ T_{\mu\nu}^F$, defined by
\begin{equation}
  { T}^G_{\mu\nu}(x)= 
  {1\over g_0^2} { F}^a_{\mu\alpha}(x){ F}^a_{\nu\alpha}(x) - 
  \delta_{\mu\nu}{{\mathcal{L}}}^G(x), \qquad \mbox{with}\qquad
{ F}_{\mu\nu}^a(x)=2\,\mbox{Tr}\{{\widehat F}_{\mu\nu}(x)T^a\}
\end{equation}
\begin{equation}
  { T}^F_{\mu\nu}(x)=
  \frac{1}{8}\Big\{
  \psibar(x)\gamma_\mu\big[\nabdbarstar\nu+\nabdbar\nu\big]\psi(x)+
  \psibar(x)\gamma_\nu\big[\nabdbarstar\mu+\nabdbar\mu\big]\psi(x)\Big\}
  -{1\over 4} \delta_{\mu\nu} {\mathcal{L}}^F(x),
\end{equation}
\begin{equation}
  {\mathcal{L}}^G(x)=
  {1\over 4g_0^2} {F}^a_{\alpha\beta}(x){F}^a_{\alpha\beta}(x),
\qquad \mbox{and}\qquad
  {\mathcal{L}}^F(x) =  
  \psibar(x)\Big\{\frac{1}{4}\gamma_\alpha\Big(\nabdbarstar\alpha+\nabdbar\alpha\Big)+M_0\Big\}\psi(x),
\end{equation}
where $\widehat F_{\mu\nu}  (x) $ is the clover discretization of the field strength tensor, $T^a$ are the generators of the gauge group
$SU(N_c)$ and the covariant derivatives are $\nabdbar\mu = \nab\mu - \nabbar\mu$ and $\nabdbarstar\mu = \nabstar\mu - \nabbarstar\mu$ with
\begin{equation}
  \nabla_\mu \psi (x) = U_\mu(x) \psi(x+\hat \mu) -\psi (x),
  \qquad
  \nabla_\mu^* \psi (x) = \psi (x) - U_\mu^\dagger (x-\hat \mu) \psi(x-\hat \mu),
\end{equation}
\begin{equation}
  \psibar(x)\nabbar\mu = \psibar(x+\hat{\mu})U^\dag_\mu(x) - \psibar(x),
  \qquad
  \psibar(x)\nabbarstar\mu = \psibar(x)-\psibar(x-\hat{\mu})U_\mu(x-\hat{\mu}).
\end{equation}
In this context, an interesting observable is the flavour-singlet conserved lattice vector current 
\begin{equation}
  V^c_\mu(x)={1\over 2}
  \Big[ \psibar(x+\hat{\mu})U^\dag_\mu(x) (1+\gamma_\mu)\psi(x) + \psibar(x)U_\mu(x) (\gamma_\mu-1)\psi(x+\hat{\mu}) \Big]
\end{equation}
whose temporal component is related to the free-energy density by the following equation 
\begin{equation}\label{Vcurrent}
  i\langle V^c_\mu(x) \rangle_{\boldsymbol \xi,\theta} = 
 L_0 {\partial \over \partial \theta} f(L_0;\boldsymbol \xi,\theta)
\end{equation}
where $\langle \cdot \rangle_{\boldsymbol \xi,\theta}$ stands for the expectation value w.r.t. the partition function of
eq.~(\ref{partfunct}). 

\section{Non-perturbative renormalization of the EMT}
In this section we discuss the conditions for the non-perturbative renormalization of the EMT on the lattice. The approach is based on
imposing for renormalized quantities on the lattice the validity of some equations -- namely Ward Identities (WIs) -- that hold in the
continuum and that follow from the fact that the EMT contains the generators of the Poincar\'e symmetry. 

The trace of $ T_{\mu\nu}$ is a singlet $\{1\}$  since it is invariant under the hypercubic group; the
off-diagonal elements transform under the $\{6\}$ representation and the diagonal elements of the traceless EMT belong to the $\{3\}$
representation. Those three parts renormalize independently and the renormalized EMT can be written as
$ T_{\mu\nu}^R= T_{\mu\nu}^{R,\{6\}}+ T_{\mu\nu}^{R,\{3\}}+ T_{\mu\nu}^{R,\{1\}}$. For a given
representation of the hypercubic group, the gluonic and the fermionic parts mix together and for the sextet and the triplet components we
have
\begin{equation}
  { T}_{\mu\nu}^{R,\{6\}}= Z_G(g_0){ T}_{\mu\nu}^{G,\{6\}}+Z_F(g_0){ T}_{\mu\nu}^{F,\{6\}}, \qquad 
  { T}_{\mu\nu}^{R,\{3\}}= z_G(g_0){ T}_{\mu\nu}^{G,\{3\}}+z_F(g_0){ T}_{\mu\nu}^{F,\{3\}}.
\end{equation}
The discussion of the renormalization of the trace is postponed to a paper we have in preparation~\cite{DallaBrida:2019}. It is important to
point out that, since the EMT in the continuum limit has no anomalous dimension, the renormalization constants are scale independent and
they depend only on the bare coupling $g_0$.

The first interesting relation is a consequence of the fact that the space-time components of the EMT are the generators of translations: it
relates the free-energy density with the one-point function of the EMT and it is given by
\begin{equation}
  \label{eq:WI1}
  \langle  T_{0k}^{R,\{6\}} (x) \rangle_{\boldsymbol \xi,\theta} = 
  -{\partial \over \partial \xi_k} f(L_0;\boldsymbol \xi,\theta)
\end{equation}
where the symbol for the derivative w.r.t. the $k$-component of the shift ${\boldsymbol \xi}$ on the r.h.s should be interpreted in terms of a
discrete approximation on the lattice. Since we have to solve that equation for the two unknown renormalization constants, $Z_G$
and $Z_F$, a first option is to consider eq.~(\ref{eq:WI1}) for two different values of the fermionic phase $\theta_1$ and
$\theta_2$. Another possibility is to use eq.~(\ref{Vcurrent}) and write
\begin{equation}
  \label{eq:WI2}
  i{\partial \over \partial \xi_k}\langle V^c_0(x)\rangle_{\boldsymbol \xi,\theta} 
  =
  - L_0 {\partial \over \partial \theta}\langle { T}^{R,[6]}_{0k}(x) \rangle_{\boldsymbol \xi,\theta},
\end{equation}
This equation can also be expressed in integral form integrating over $\theta$. We emphasize that, since the Wilson discretization preserves
the baryonic symmetry of the continuum theory, the vector current $V^c_\mu$ is exactly conserved on the lattice and it does not renormalize.

Once the renormalization constants of the sextet representation are fixed, the renormalization constants of the triplet representation can
be obtained by requiring in the thermodynamic limit that
\begin{equation}
  \label{eq:WI3a}
  \langle { T}^{R,\{6\}}_{0k}(x) \rangle_{\boldsymbol \xi,\theta}={\xi_k\over 1-\xi_k^2}\;
  \langle { T}^{R,\{3\}}_{00}(x) -
          { T}^{R,\{3\}}_{kk}(x) \rangle_{\boldsymbol \xi,\theta} ,
  \qquad
  \theta=\theta_1,\theta_2
\end{equation}
or else
\begin{equation}
  \label{eq:WI3b}
  \langle { T}^{R,\{6\}}_{0k}(x) \rangle_{\boldsymbol \xi,\theta}=\xi_k \;
  \langle { T}^{R,\{3\}}_{00}(x) -
          { T}^{R,\{3\}}_{jj}(x) \rangle_{\boldsymbol \xi,\theta} ,
  \qquad
  \theta=\theta_1,\theta_2
\end{equation}
with $j\neq k$ and $\xi_j=0$. The above two WIs follow from applying to the EMT the Lorentz transformation that changes the rest frame to
the moving frame described by the shifted boundary conditions~\cite{Giusti:2012yj}.
The WIs that we have discussed above provide a non-perturbative definition of the renormalization constants of the EMT on the
lattice. Interestingly, they also represent a practical method for computing those constants by Monte Carlo simulations on the lattice
similar to the case of the Yang-Mills theory~\cite{Giusti:2015daa}. Work is in progress in this respect. Finally, the validity of the above
proposal has been checked by a calculation at 1-loop order in perturbation theory~\cite{DallaBrida:2019}.

\section{The QCD Equation of State}

The integral method is the technique usually considered for studying the Equation of State of a quantum theory and it is based on
the direct measurement of the trace anomaly of the EMT~\cite{Boyd:1996bx}. The pressure, the entropy density and the energy density can then
be obtained using standard thermodynamic relations. Although that technique has been effective and very
successful, it requires to take into account a second temperature scale other than the temperature at which one is interested in. In fact, the
operator whose expectation value gives the trace anomaly, mixes with the identity with the consequence of having ultraviolet power
divergences. The difference of the trace anomaly at two different temperatures cancels those divergences but, from the numerical
viewpoint, it can be a challenge since one has to perform Monte Carlo simulations with the same bare parameters at two different
temperature scales. These computational difficulties have allowed to investigate the QCD Equation of State in the continuum limit only up to
temperatures of about 1 GeV and between 1 GeV and 2 GeV only with a quite large value of the lattice spacing~\cite{Borsanyi:2013bia,
  Bazavov:2014pvz, Bazavov:2017dsy}. Furthermore, those results have been obtained by considering the rooted staggered formulation of the fermionic
fields on the lattice: it is a convenient choice on the numerical side but not fully satisfactory from the theoretical viewpoint.

The framework of shifted boundary conditions provides a much simpler environment to investigate the Equation of State. In that setup the
primary observable that is measured in Monte Carlo simulations is the entropy density, $s(T)$: it is directly related to the expectation
value of the space-time components of the renormalized EMT by the equation
\begin{equation}\label{entden}
  {s(T)\over T^3} = - {L_0^4 (1+\boldsymbol \xi^2)^3 \over \xi_k} 
  \langle { T}^{R}_{0k}(x) \rangle_{\boldsymbol \xi,0}=
  - {L_0^4 (1+\boldsymbol \xi^2)^3 \over \xi_k} \Big[
  Z_G \langle { T}_{0k}^{G,\{6\}}\rangle_{\boldsymbol \xi,0}
  +Z_F \langle { T}_{0k}^{F,\{6\}}\rangle_{\boldsymbol \xi,0} \Big].
\end{equation}
There are no divergences to remove and, once the renormalization constants are known, one just needs to perform Monte Carlo simulations to compute
the two expectation values that appear on the r.h.s. The continuum limit is attained in a simple way just by tuning the bare parameters so
to run the numerical simulations at the physical temperature of interest. Furthermore the numerical challenges that prevent the integral
method to increase the temperature above 1-2 GeV are overcome since one does not need to take into account two temperature scales at the
same bare parameters. The framework of shifted boundary conditions turned out to be efficient and successful in investigating the Equation
of State of the $SU(3)$ Yang-Mills theory with a permille accuracy up to very high temperatures~\cite{Giusti:2014ila,Giusti:2016iqr}. We now
discuss its application to QCD in the completely unexplored range of temperatures between 2 GeV and 80 GeV, i.e. the electro-weak scale. We also
point out that for the pure gauge theory, it turns out that high temperature perturbation theory does not represent a reliable description of
the gluon plasma and it would be interesting to study what happens for QCD. 

We perform numerical simulations of lattice QCD with $O(a)$-improved Wilson fermions with the standard plaquette Wilson action for the gauge
sector. We consider 3 flavours of degenerate massless quarks: this choice is justified by the fact that we plan to investigate a regime of
temperatures where the effects related to the quark masses can be neglected. We study the Equation of State at 8 physical
temperatures $T$ equally spaced in log scale~\cite{DallaBrida:2017sxr,DallaBrida:2019gdt}. For each temperature we consider 4
values of the lattice spacing corresponding to $L_0 =$ 4, 6, 8 and 10. The system size in the spatial directions is kept fixed at
$L =$ 288 and we consider the shift $\boldsymbol \xi = (1,0,0)$: thus, $LT$ ranges from 20 to 51 and finite size effects can be
neglected. The numerical simulations are almost complete: only the last runs for 3 temperatures with $L_0 =$ 10 are in progress. The values of
the bare coupling have been determined using the results of the ALPHA collaboration for the running
coupling~\cite{Brida:2016flw,DallaBrida:2016kgh,Bruno:2017gxd,DallaBrida:2018rfy,Campos:2018ahf}. 

Finally, in order to obtain the measurement for the entropy density in eq.~(\ref{entden}) and to perform the extrapolation to the continuum
limit we need to compute the sextet renormalization factors, $Z_G$ and $Z_F$: this work is in progress.

{\bf Acknowledgements}
We acknowledge PRACE for awarding us access to MareNostrum at Barcelona Supercomputing Center (BSC), Spain (n. 2018194651). We also
acknowledge support from a CINECA-INFN agreement, providing access to Marconi at CINECA, Italy.

\end{document}